\documentclass[aps,prd,nofootinbib,preprintnumbers,twocolumn,showpacs, floatfix]{revtex4-1}
\usepackage{amsmath}
\usepackage{amssymb}         
\usepackage{epsfig}
\usepackage{graphicx}
\usepackage{multirow}
\usepackage{color}
\usepackage[normal]{subfigure}
\usepackage{rotating}
\usepackage{hyperref}
\usepackage{textcomp}

\newcommand{\ewt}{\end{widetext}}
\newcommand{\be}{\begin{equation}}
\newcommand{\ee}{\end{equation}}
 \newcommand{\bdm}{\begin{displaymath}}
\newcommand{\edm}{\end{displaymath}}
\newcommand{\bea}{\begin{eqnarray}}
\newcommand{\eea}{\end{eqnarray}}

\begin{document}
\title{ Dilepton events with displaced vertices, double beta decay,
  and resonant leptogenesis with Type-II seesaw dominance, TeV  scale $Z'$ and
  heavy neutrinos}
\author{ Bidyut Prava Nayak and M. K. Parida}\email{bidyutprava25@gmail.com}\email{minaparida@soauniversity.ac.in}

\affiliation{Centre of Excellence in Theoretical and Mathematical sciences\\
SOA University, Khandagiri Square, Bhubaneswar 751030, India}

\begin{abstract}
 In a class of Type-II seesaw dominated $SO(10)$ models proposed recently with 
heavy neutrinos, extra $Z'$ boson, and resonant leptogenesis, at first
we show that the lightest first generation
 sterile neutrino that mediates dominant contributions to neutrinoless  double beta decay also generates the
  displaced vertex leading to verifiable  
 like-sign di-electron as well as  di-muon production events outside
 the LHC detectors having
 suppressed standard model back ground and missing energy. Resonant leptogenesis in
 this case is implemented by a pair of quasi-degenerate sterile
 neutrinos of 
the second
 and the third generations having  masses of ${\cal O}(500)$ GeV. Then we
 predict a new alternative scenario where the models allow
 the second generation
 sterile neutrino mass to be ${\cal O}(10)$ GeV capable of mediating the
 dominant double beta decay as well as the displaced vertices for 
 significantly improved number of
 like-sign dilepton events in different channels.
 Resonant
 leptogenesis in this alternative scenario
  is mediated by a pair of heavy quasi-degenerate sterile
 neutrino masses  of the first and the third
 generations. In addition to QD type light neutrino mass hierarchy ,
 we also show how these results are
 derived for the NH case.
We also discuss $Z'$ production cross
 sections at $\sqrt s =14 $ TeV  run-II of LHC and also at ILC.  
While the lepton flavour violating branching ratios are
 only few to four orders less than the current experimental bounds,
  proton lifetime predictions are accessible to ongoing Super K. or
 Hyper K. searches. 
\end{abstract}
\maketitle

\section{Introduction}

The renormalisable standard model (SM) predicts the neutrinos to be massless 
 whereas neutrino oscillation experiments prove them
to be massive \footnote{The
  nonrenormalisable $\text {dim}.5$ operator scaled by the Planck mass
  gives too small a Majorana neutrino mass $\sim {10}^{-5}$ eV \cite{Weinberg:1979} to
  account for the oscillation data.}. Theoretically, these masses of the light neutrinos are predicted through
various seesaw
mechanisms \cite{rev:2006,type-I,shafi,type-III,ma:2006,mkp:2011,LG:2007} such as
Type-I, Type-II, Type-III, double seesaw, inverse seesaw \cite{vallernm:1986}, and
radiative seesaw mechanisms in SM extensions. In a minimal left-right
symmetric \cite{rnmpati:1975,pati:1973} grand unified theory (GUT)
like $SO(10)$ \cite{georgi:1974} where the
origin of parity (P) violation in weak interaction is explained \cite{baburnm:1993}, a number
of these seesaw mechanisms can be naturally embedded while answering
the origin of the three gauge couplings of the SM. With natural inclusion
of right-handed (RH) neutrinos in its spinorial fermionic representation, the
model can predict the Dirac neutrino masses or Yukawa couplings by
fitting the charged fermion masses which play a crucial role in
 seesaw mechanisms, lepton flavour violations (LFVs), and lepton number
violations (LNVs).
The GUT model has also the potential to explain baryon asymmetry of the universe via
leptogenesis through heavy RH neutrino decays \cite{fukuyana:1986}.
Although recently several attempts have been made to search for smoking gun signatures of TeV scale left-right gauge theory from available LHC measurements at $\sqrt s=8$ TeV, no conclusion can  be reached due to poor statistics of the available data and until different sources of standard model background events are clearly identified. 
 
However it is possible that the associated intermediate scale of LR  gauge theory in GUTs and the $W_R$ boson mass could be too large to
be accessible to accelerator tests in near future and to manifest in
ongoing experimental searches for
$0\nu\beta\beta$ decay \cite{bbexptz,bbexpt1,bbexpt2,bbexpt3}. Inspite of this, its neutral $Z^{\prime}$  gauge boson as a smoking gun signal of underlying high scale left-right gauge symmetry 
and the associated RH neutrinos could be near the TeV scale \cite{mkpjcp,zpmod1,montero:2010,Accomando,Andreev,carena,ramlalmkp,app,pas,bpnmkp:2015}. Basically
this scheme materialises in two-step breaking of the LR gauge theory, 
 \\
\par\noindent
$SU(2)_L\times SU(2)_R\times U(1)_{B-L}\times SU(3)_C (=G_{2213},g_{2L}\neq g_{2R})$\\
\begin{center}
 OR
\end{center}
\par\noindent
$SU(2)_L\times SU(2)_R\times U(1)_{B-L}\times SU(3)_C (=G_{2213D},g_{2L}= g_{2R})$\\
\bea
& \longrightarrow &
SU(2)_L\times U(1)_R\times U(1)_{B-L}\times SU(3)_C (=G_{2113})\nonumber\\
&&\nonumber\\
&\longrightarrow& SM \label{lro}
\eea
Alternatively $G_{2113}$ can emerge from direct breaking of Pati-Salam gauge theory  $SU(2)_L\times SU(2)_R\times SU(4)_C$ or a GUT like $SO(10)$, or $E_6$.
In a recent paper\cite{bpnmkp:2015} we embedded this two-step symmetry
breaking chain in two models originating from 
nonsupersymmetric (non-SUSY) $SO(10)$ \\ 

\par\noindent{\bf Model-I}
\bea
SO(10)\to G_{2213D} \to G_{2113} \to SM. \label{mod-1}
\eea 
\par\noindent{\bf Model-II}
\bea
SO(10)\to G_{2213} \to G_{2113} \to SM. \label{mod-2}
\eea 
In Model-I, the first step of symmetry breaking takes place by
assigning the GUT-scale VEV to the neutral component of the Higgs
submultiplet $(1,1,15)\subset {210}_H$ of $SO(10)$ under Pati-Salam
gauge symmetry $SU(2)_L\times SU(2)_R\times SU(4)_C$. As this neutral
component carries D-Parity even quantum number \cite{cmp}, the GUT symmetry breaks without breaking D-Parity. In Model-II, the D-Parity itself 
breaks down at the GUT scale by assigning, in addition, the GUT scale
VEV to the D-Parity odd singlet component $(1,1,1)_H\subset {210}_H$ \cite{cmp}. The second step
of symmetry breaking in both models is implemented  by assigning 
intermediate scale VEV to the Higgs scalar component $\sigma(1,3,0, 1)\subset
 {45}_H $. The third step of symmetry breaking in both models is
 materialised by assigning TeV scale VEV to the
neutral component of the RH scalar triplet $\Delta_R (1,3,-2,1)$ which
generates the TeV scale $Z'$- boson and the RH neutrino masses. 
The Type-II seesaw dominance occurs in Model-I by the natural presence of  
the LH
triplet $\Delta_L(3,1,-2,1)\subset {126}_H$  that acquires the desired
induced VEV  needed to drive the seesaw mechanism. Type-II seesaw
contribution dominates by suppressing the linear seesaw term by
appropriate finetuning of parameters. In Model-II the mass of the LH
triplet  $\Delta_L(3,1,-2,1)$ is kept at the intermediate scale by
fine tuning of parameters to implement type-II seesaw seesaw dominance
while the linear seesaw term is naturally suppressed in this case.

The $W_R$ boson mass being $> 10^8$ GeV in both models,  at first sight it appears that there are no additional contributions
to  $0\nu\beta\beta$ decay other than the standard contributions due
to light neutrinos though their NH, IH, or QD patterns of tiny
Majorana masses and the well known structure of the PMNS mixing matrix.
But the models of eq.(\ref{mod-1}) and eq.(\ref{mod-2}) have been specifically designed to
include additional non-standard fermion singlets of three generations which acquire
Majorana masses to mediate dominant contributions to $0\nu\beta\beta$ decay
irrespective of the hierarchy of light neutrino masses.

 The other two of
the sterile neutrinos being quasi-degenerate with masses around ${\cal O}(1)$ TeV have
been shown \cite{bpnmkp:2015} to mediate resonant leptogenesis \cite{pilaftsis:2003} explaining baryon
asymmetry of the Universe. These heavy Majorana neutrinos including
the RH and sterile ones may also manifest in the production of like-sign
dilepton signals at ATLAS or of the type observed at  CMS detectors \cite{CMSMN}  provided
adequate beam luminosity is reached.

Quite recently, as a very interesting and novel manifestation of
Majorana type of RH neutrino that occurs in  Type-I seesaw mechanism, it has been pointed out that if $W_R$ boson is at the
TeV scale  and a RH neutrino mass needed for the seesaw is sufficiently
light, 
it would mediate  $0\nu\beta\beta$ decay while like-sign di-electron signals
caused due to displaced vertices mediated by the RH neutrino
mass in the range
$1-80$ GeV would provide more interesting model signatures through
$eejj$ events devoid of
standard model back grounds \cite{Helo,Kovel} without missing
energy. Then these like-sign
di-electron signals and 
$0\nu\beta\beta$ decay events would indicate the presence of
the gauge theory at the TeV scale. Even if there is no $W_R$ gauge
boson at the TeV scale, this approach predicts the novel possibility of
like-sign dilepton events outside the ATLAS or CMS detectors with
suppressed SM back ground and missing energy in the channel $eejj$ in
the SM extension with a RH neutrino to accommodate Type-I seesaw.
 In such a Type-I seesaw
model as the associated single RH neutrino is sufficiently light, it
is difficult to implement TeV scale resonant leptogenesis for which two
quasi-degenerate heavy masses of RH neutrinos of two other generations
may be needed and this needs further investigation. Also such a
single RH neutrino model may not adequately
mediate inside detector events at CMS or ATLAS in the channels $pp\to \mu\mu
jj X$, or $pp \to e\mu jj X$. In addition, the heavy-light neutrino
mixings in this model can be any where bounded by the DELPHI
\cite{DELPHI} and
the  double beta decay experimental limits.  

The purpose of this work is two fold:(i) In the first part we show that
the two non-SUSY $SO(10)$ models discussed recently
\cite{bpnmkp:2015}, as indicated by eq.(\ref{mod-1}) and
eq.(\ref{mod-2}), predict a rich structure of like-sign di-electron and
di-muon events with displaced vertices outside the LHC detectors along with dominant
contributions to double beta decay mediated by lighter masses of
one of the three sterile neutrinos while TeV scale masses of Majorana
neutrinos are available to mediate inside detector events like
$pp\to l^{\pm}l^{\pm}jj X$. In contrast to models 
along this line including those of ref.\cite{Helo,Kovel} where heavy
-light neutrino  mixings are assumed under the DELPHI \cite{DELPHI}
and the double beta decay constraints , these
 mixings in our models are predicted from all charged fermion mass fit at
 the GUT scale and  the LFV constraint. The Dirac neutrino mass matrix  derived in this manner serves
as important ingredient for predictions of LFV, LNV, and dilepton
production events. In addition, our models provide a mechanism
for resonant leptogenesis \cite{pilaftsis:2003} mediated by TeV scale quasi-degenerate pair
of sterile neutrinos. Two different cases have been identified. In the Case (a) discussed in this paper while
the light sterile neutrino $S_1$ of first generation mediates dominant
contribution to double beta decay and like-sign dilepton events with
displaced vertices in the channels $eejj$, $e\mu jj$, and $\mu\mu jj$,
resonant leptogenesis is allowed to be mediated by heavy
quasi-degenerate sterile neutrino pairs $S_2$ ans $S_3$ belonging to
the second and the third generations, Identified as the alternative
Case (b), dominant double beta decay and dilepton events with displaced
vertices are mediated by the allowed lighter mass of $S_2$ while resonant
leptogenesis is mediated by the heavy quasi-degenerate sterile
neutrino pairs, $S_1$ and $S_3$. 
 In addition to QD
hierarchy of light neutrino masses, we also show how all these results
hold in the presence of NH masses, a result which might be important
if the recent cosmological bound \cite{cosnu:2014} is finally established. (ii) In the second part of this work we explore
detection possibilities of the extra $Z'$ boson predicted by these two
models at the Large Hadron Collider (LHC) and the International Linear
Collider (ILC). We also predict heavy RH Majorana neutrino production cross
sections through the like-sign dilepton production at the LHC
detectors in the $W_L-W_L$ channel.     

This paper is organized in the following manner.
In sec.2 we provide a brief description of the model results
investigated earlier. In sec.3 we discuss the neutrinoless double
 beta decay by sterile neutrino exchange. In sec.4 we discuss
 resonant leptogenesis in the cases where
the first or the second generation sterile neutrino is light. 
In sec.5 we discuss decay width, half-life and displaced length
due to the mediation of light sterile neutrinos
in the model. In sec.6 we discuss  the predictions of cross sections
and observable number of events due to  displaced vertices.
The heavy RH neutrino mediated dilepton production cross section in the $W_L-W_L$ channel is discussed in sec.7
In sec.8 we
discuss the neutrinoless double beta decay and leptogenesis with NH neutrinos.
The $Z'-$ boson production cross section and its comparison with the
standard $Z-$ boson cross section are discussed in sec.9.
 In sec.10  we summarize this work and draw conclusion.\\

\section{\bf TYPE-II SEESAW DOMINANCE IN $SO(10)$}
Several interesting approaches have been made earlier to implement type-II
seesaw dominance for neutrino masses in $SO(10)$ \cite{nasri:2004,Melfo:2010,rnmmkp11}. 
As discussed below while giving a brief summary of common aspects of  Model-I and
Model-II \cite{bpnmkp:2015} in non-SUSY $SO(10)$ GUT relevant to the
present work, we note that in both the models excellent gauge coupling
unification has been achieved with proton lifetime predictions over a
wider range of values in each model covering the experimentally
accessible search limits \cite{babu:pdecay,perez:2007} for 
$p \to e^+\pi^0$ mode.

\subsection{\bf Light and Heavy Neutrino Masses}

The Yukawa Lagrangian of the two  models \cite{bpnmkp:2015} is given by
\bea
\mathcal{L}_{\rm Yuk} &=&  \sum_{i=1,2}Y_{i}^{\ell} \overline{\psi}_L\, \psi_R\, \Phi_{i} 
                       + f\, (\psi^c_R\, \psi_R \Delta_R +\psi^c_L\, \psi_L \Delta_L)\nonumber\\
&&+y_{\chi}\, (\overline{\psi}_R\, S\, \chi_R
                       +\overline{\psi}_L\, S\, \chi_L) \nonumber\\  
                       &&+{\mu}_S SS + \text{h.c.},
                       \label{Yuk-Lag} 
                       \eea 
where ${\mu}_S$ represents the global lepton number violating mass
term which is naturally and vanishingly small in the t'Hooft sense \cite{tHooft}.
  
                      Using the VEVs of the Higgs fields and denoting 
$M_N=f<\Delta_R>=fV_R$, $M=y_{\chi}<\chi_R> =y_{\chi}V_{\chi}$,   
$M_D = Y_i<{\Phi}_i> + X$, where the origin of
the term $X$ has been discussed earlier \cite{bpnmkp:2015}, 
a $9 \times 9 $ neutral-fermion mass matrix has been obtained which, upon
block diagonalisation, yields  $3\times 3$ mass
matrices each for the light neutrino ($\nu_{\alpha}$), the right handed
 neutrino ($N_{\alpha}$), and the sterile neutrino
 ($S_{\alpha}$) . Including the induced Type-II
 seesaw contribution in both the models, the following generalised
 form of the $9\times 9$ matrix has been obtained in the
$(\nu, S, N)$ basis

\bea
\mathcal{M}=
\begin{pmatrix}
m_{\nu}^{II}    &   M_L     &  M_D  \\
M_L^T &   \mu_S      &  M  \\
M_D^T &   M^T     &  M_N
\end{pmatrix},
\label{app:numass}
\eea 
where $m_{\nu}^{II}=fv_L$, $v_L$ being the induced VEV of the neutral
component in the LH triplet 
$\Delta_L (3, 1, -2, 1) \subset {\overline {126}}_H$.   
On block diagonalisation under the extended seesaw constraint,
$M_N > M > M_{D},\mu$, it has been shown that the Type-I seesaw term
cancels out \cite{bpnmkp:2015,Kimkang:2006,Majee:2007} and the generalised form of the light neutrino mass matrix is given by
\bea
{\cal M}_{\nu} &=&~ fv_L+M_LM^{-1}M_N(M^T)^{-1}M_L^T \nonumber\\
&&-[M_LM_D^TM^{-1} +M_DM_L^T{M^T}^{-1}]\nonumber\\
&&+ (M_D/M)\mu_S {(M_D/M)}^T ,\label{nugen}
\eea
Since the LH doublet $\chi_L$ has vanishing VEV as explained in \cite{bpnmkp:2015}, the $M_L$ term drops out and the
Type-II seesaw dominance prevails in the limit of $^{'}$t Hooft's
naturalness condition,  $\mu_S \to 0$ \cite{tHooft}
 
\be
{\cal M}_{\nu} \simeq fv_L.\label{T2}
\ee
  With  Type-II seesaw dominated neutrino mass
formula,
the RH neutrino masses are also predicted in this scenario.
\bea
M_N=fV_R={\cal M}_{\nu}\frac{V_R}{v_L}.\label{matMN} 
\eea 

At first we will discuss the case when light neutrino masses are
quasi-degenerate (QD), subsequently we will also show how our results are  applicable for normally hierarchical (NH) case. 
 As reported \cite{bpnmkp:2015} with
 $m_{\nu_1}= 0.2056$ eV, $m_{\nu_2}= 0.2058$ eV and $m_{\nu_3}= 0.2$ eV, and
$v_L=0.5$ eV, the neutrino mass matrix, the Yukawa matrix $f$, and the
 RH neutrino mass eigen values are  

\bea
 m_{\nu} =
\begin{pmatrix} 1.01+0.01i & 0.00055+0.01i  &  -0.009+0.1i\\
0.00055+0.01i  & 1.01+0.008i  & 0.01+0.105i\\-0.009+0.1i &0.01+0.1i&
0.9-0.02i \end{pmatrix}~{\rm eV} \nonumber\\
\label{fQD}
\eea
\bea
 f =
\begin{pmatrix} 2.02+0.02i & 0.0011+0.02i  &  -0.019+0.3i\\
0.0011+0.02i  & 2.034+0.017i  & 0.021+0.21i\\-0.019+0.3i &0.021+0.21i&
1.99-0.04i \end{pmatrix}\nonumber\\
 \label{fQD}
\eea
\bea
 |{\hat M}_N| ={\rm diag}( 4.08, 4.03, 4.02) ~{\rm TeV}.\ \label{MNQDp5}
\eea

The mass matrix of the sterile neutrino is 
\bea
m_S=-M\frac{1}{M_N}M^T \label{matms}
\eea
where $M$ is the $N-S$ mixing mass term in the Yukawa Lagrangian of eq.(\ref{Yuk-Lag}). 

The Dirac neutrino mass matrix plays crucial roles in predicting lepton flavour
 violating branching ratios, leptonic non-unitarity effects, lepton
 number violations, leading to heavy-light neutrino mixings which act as important ingredients
in estimating the model predictions for di-lepton events at LHC as discussed
below. In both our models this mass matrix has been determined by fitting the 
RG extrapolated values of charged fermion masses at the GUT scale \cite{dp:2000} and extrapolating it back to the lower scale $\mu\sim M_{Z'}\sim M_{R^0}$ 

\bea
M_D({\rm GeV})= 
\footnotesize\begin{pmatrix}
0.014&0.04-0.01i&0.109-0.3i\\
0.04+0.01i&0.35&2.6+0.0007i\\
0.1+0.3i&2.6-0.0007i&79.20
\end{pmatrix}.\nonumber\\
\label{MD} 
\eea 
This procedure has been followed in a number of recent works in
non-SUSY $SO(10)$ with TeV scale $Z'$ and RH Majorana neutrinos
\cite{app} and in SUSY SO(10) with TeV scale $W_R$ but pseudo-
Dirac neutrinos \cite{psbrnm}. In the non-SUSY $SO(10)$ model with TeV
scale $Z'$ and heavy RH pseudo-Dirac neutrinos, the corresponding matrix
has been derived in ref.\cite{ramlalmkp}.
      
\subsection{\bf Lepton Flavour Violations}

In all conventional non-SUSY $SO(10)$ GUTs with high scale 
Type-I or Type-II seesaw formula for neutrino masses, there are
three generations of standard fermions in ${\bf {16}}_i(i=1,2,3)$ and
the RH neutrino masses are large which give negligible contribution to
charged LFV decay amplitudes. In the case of non-SUSY SO(10) with 
inverse seesaw mechanism
for neutrino masses, the three singlet fermions \footnote{ Each of these three
  fermion singlets may belong to the non-standard fermionic
  representation ${45}_F\subset SO(10)$ or the fundamental
  representation ${27}_F\subset E_6$.} with their mixing mass term $M$
near the TeV scale and
through their loop mediation make substantial contribution to LFV
decay branching ratios accessible to ongoing experimental searches \cite{ramlalmkp}.
In the present non-SUSY SO(10) model, even though neutrino masses are
governed by high scale type-II seesaw formula, the essential presence of
singlet fermions $S_{i}$ that implement the Type-II seesaw dominance by
cancelling out the Type-I seesaw contribution give rise to
experimentally observable LFV decay branching ratios through their
$N-S$ mixing mass terms in loop mediation.
 The heavier RH neutrinos in this model being in the
range of $\sim 0.1-1$ TeV  also contribute, but less
significantly than the singlet fermions. The predicted branching ratios 
being only few to four orders less than the current experimental
limits are verifiable by ongoing searches \cite{adam},

\bea
BR(\mu \to e\gamma)&=&6.43\times 10^{-17},\nonumber\\
BR(\tau \to e\gamma)&=&8.0\times 10^{-16},\nonumber\\
BR(\tau \to \mu\gamma)&=&2.41\times 10^{-12}.\label{lfvbr}
\eea  
  
\section{\small\bf HEAVY-LIGHT MIXING AND PREDICTIONS FOR NEW PHYSICS
  EFFECTS}
In ref. \cite{bpnmkp:2015} we have discussed how the lightest sterile
neutrino of first generation gives rise to dominant double beta decay
and the heavier quasi-degenerate pair of second and the third generation
sterile neutrinos produce baryon asymmetry via resonant leptogenesis \cite{pilaftsis:2003}.
In this section we point out that, in addition to leptogenesis, the models have much wider impact on
lepton number violating phenomena to be detectable by ongoing double beta decay
experiments at low energies and like-sign dilepton production events
at LHC via displaced vertices. In  predicting the LFV
branching ratios we have used the simplifying diagonal structure for
M,
\be
M = {\rm diag.}~(M_1, M_2, M_3),\label{EQM}
\ee
which in combination with eq.(\ref{MD}) gives the elements of the
  $\nu_l-S$ mixing matrix 
\bea
{\cal V}^{(lS)}=
\begin{pmatrix}
{M_D}_{e1}/M_1& {M_D}_{e2}/M_2&{M_D}_{e3}/M_3\\ 
{M_D}_{\mu 1}/M_1& {M_D}_{\mu 2}/M_2&{M_D}_{\mu 3}/M_3\\ 
{M_D}_{\tau 1}/M_1& {M_D}_{\tau 2}/M_2&{M_D}_{\tau 3}/M_3
\end{pmatrix}.
\label{HLMIX}
\eea
Only the physical manifestation of ${\cal V}_{e1}^{(lS)}$ through
new dominant prediction in double beta decay mediated by $S_1$ was discussed in
\cite{bpnmkp:2015}. But here we show that through this mixing
element, $S_1$ would also be able to mediate displaced vertex for
like-sign di-electron production events out side the LHC detector.
Interestingly, because of the significant values of the element 
 ${\cal V}_{\mu 1}^{(lS)}$, the  sterile neutrino $S_1$ will also be
able to mediate like-sign di-muon production events outside the
LHC detector.
 Similarly, in an alternative scenario, 
the presence of the mixing matrix element  ${\cal V}_{e2}^{(lS)}$
would enable $S_2$ to mediate dominant double beta decay and displaced vertex
for like-sign di-electron events outside the LHC detectors while
through the
presence of the sizeable element  ${\cal V}_{\mu2}^{(lS)}$, it would
mediate like-sign di-muon signal events via displaced vertex outside
the LHC detector. These are possible provided $S_2$ is sufficiently
light. Although similar contributions
due to the exchanges of other heavy Majorana neutrinos are possible,
they are neglected as the corresponding masses are much heavier than
${\hat m}_{S_1}$ or ${\hat m}_{S_2}$. Details of applications of these
possibilities have been discussed in the following sections. We
categorize the mediation by light $S_2$ in double beta decay and
observable dilepton production events by displaced vertices as a
second scenario, alternate to the first example where $S_1$ is light because
the resonant leptogenesis constraint requires either $S_2$ and $S_3$
in the first case, or $S_1$ and $S_3$ in the second case are to be
quasi-degenerate with mass $\sim 500$ GeV.      

\subsection{\bf Neutrinoless Double Beta Decay by Sterile Neutrino Exchange} 

\begin{figure}[htbp]
 \includegraphics[width=6cm,height=4cm]{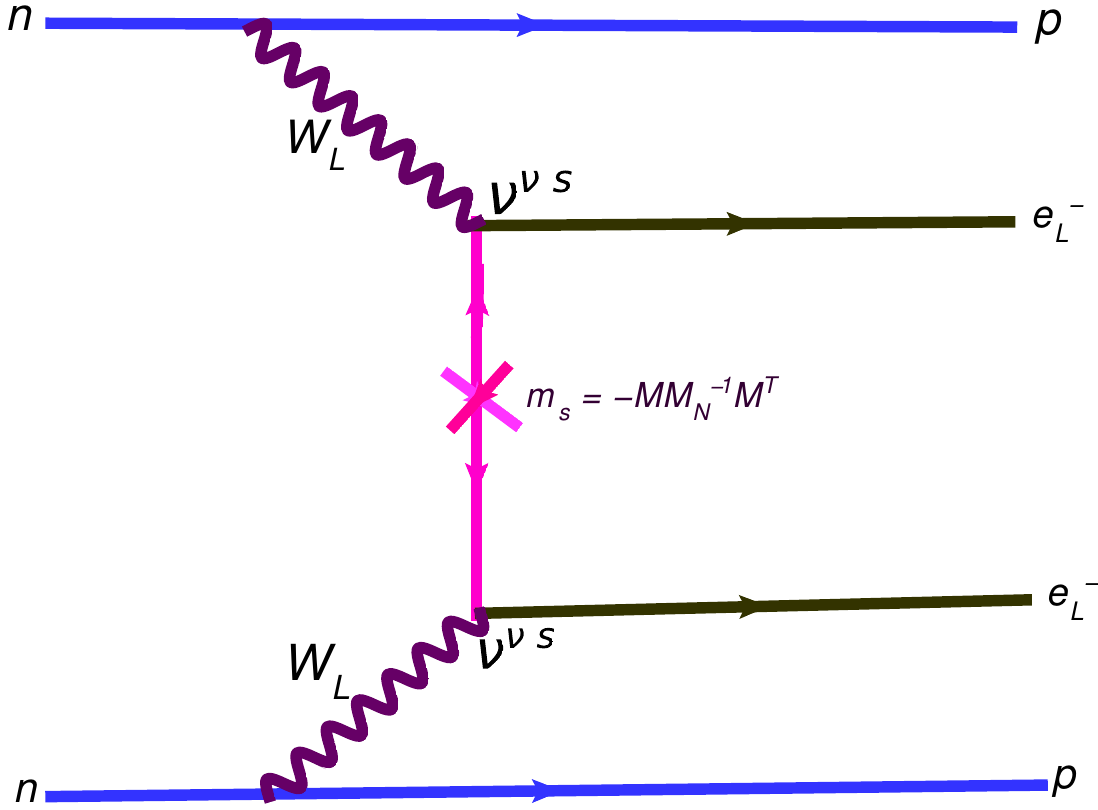}
 \caption{Neutrino-less double beta decay due to the exchange of
   sterile neutrino of first generation $S_1$ or second generation $S_2$.}   
\label{feyn1}
 \end{figure}

\begin{figure}[htb!]
\includegraphics[width=8cm,height=9cm,angle=0]{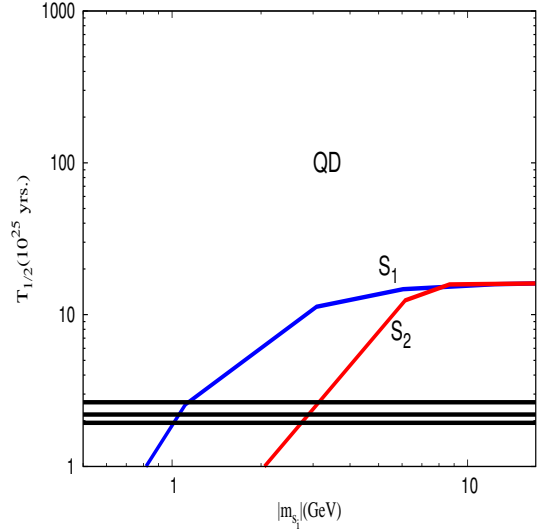}
\caption{Variation of half-life of neutrinoless double beta decay with
  sterile neutrino mass of first generation (Curve $S_1$) and second
  generation  (Curve $S_2$). The horizontal lines represent current
  experimental bounds \cite{bbexptz,bbexpt1,bbexpt2,bbexpt3}.}
\label{fig:hlfqdn}
\end{figure}

As the $W_R$ boson and the doubly charged Higgs bosons
have masses $> 10^8$ GeV, they have negligible contributions to
$0\nu\beta\beta$ decay amplitude in these models.
The Feynman diagram for sterile neutrino contribution to
$0\nu\beta\beta$ decay amplitude is shown in Fig. \ref{feyn1}.
Using suitable normalisations\cite{Vergados,Doi,Barry:2013}, the added contributions due to
 light-neutrinos, sterile neutrinos, and the heavy
RH neutrinos  in the $W_L-W_L$ channel 
lead to the inverse half life   
\begin{eqnarray}
  \left[T_{1/2}^{0\nu}\right]^{-1} &\simeq &
  G_{01}|\frac{{\cal M}^{0\nu}_\nu}{m_e}|^2|({\large\bf
    m}^{ee}_{\nu} +{\large\bf m}^{ee}_{S}+{\large\bf m}^{ee}_{N})|^2,\nonumber\\
&=& K_{0\nu}|({\large\bf
    m}^{ee}_{\nu} +{\large\bf m}^{ee}_{S}+{\large\bf m}^{ee}_{N})|^2,\nonumber\\
&=& K_{0\nu}|{\large\bf
    m}_{\rm eff} |^2
  \label{invhalf}
\end{eqnarray}
where  $G_{01}= 0.686\times 10^{-14} {\rm yrs}^{-1}$, ${\cal
  M}^{0\nu}_{\nu} = 2.58-6.64$, and $K_{0\nu}= 1.57\times 10^{-25} {\rm yrs}^{-1}
{\rm eV}^{-2}$. In eq.(\ref{invhalf}) the three effective mass parameters  are 
\begin{eqnarray}
{\large \bf  m}^{\rm ee}_{\nu} =\sum^{}_{i} \left(\mathcal{V}^{\nu \nu}_{e\,i}\right)^2\, {m_{\nu_i}}
\label{effmassparanus} 
\end{eqnarray}
\bea
{\large \bf  m}^{\rm ee}_{S} = \sum^{}_{i} \left(\mathcal{V}^{\nu
  S}_{e\,i}\right)^2\, \frac{|p|^2}{{\hat m}_{S_i}} 
  \label{effmassparanus2} 
\eea
\begin{eqnarray}
{\large \bf  m}^{\rm ee}_{N} = \sum^{}_{i} \left(\mathcal{V}^{\nu
  N}_{e\,i}\right)^2\, \frac{|p|^2}{m_{N_i}},
\label{effmassparanus3} 
\end{eqnarray}
with 
\begin{eqnarray}
&&{\large\bf m}_{\rm eff}={\large\bf m}^{ee}_{\nu} +{\large\bf
    m}^{ee}_{S}+{\large\bf m}^{ee}_{N}.\label{sumeff}
\end{eqnarray}
Here ${\hat m}_{S_i}$ is the i-th eigen value of the  $S-$ fermion mass matrix
$m_S$, and the magnitude of neutrino virtuality momentum
$|p|= 120$ MeV$-200$ MeV.
The RH neutrinos being much heavier than ${\hat m}_{S_1}$ in the first
case, or ${\hat m}_{S_2}$
in  the second case, their
contributions have been neglected. 
The variation of half-life with the first generation and second
generation sterile neutrino mass is given in  Fig. \ref{fig:hlfqdn}.

From  Fig. \ref{fig:hlfqdn} we find that that the first (second)
generation sterile neutrino saturates the current experimental
bounds at ${\hat m}_{S_1} \sim 1$ ( ${\hat m}_{S_2}\sim 2-3$)
GeV. For larger values of these mass eigen values the predicted half-life
increases but for very large values of these masses the curves would
asymptotically approach the horizontal lines as happens in the case of
QD mass hierarchy of light neutrinos when all other contributions are neglected.

\section {\bf TWO ALTERNATIVE CASES FOR RESONANT LEPTOGENESIS}

\begin{figure}[htbp]
\begin{center}
\includegraphics[width=4 cm,height=3cm]{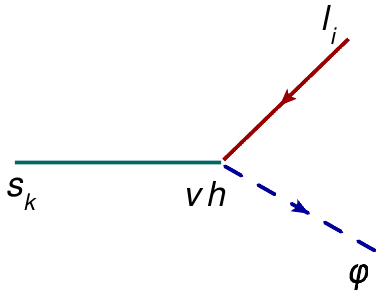}
\includegraphics[width=4 cm,height=3 cm]{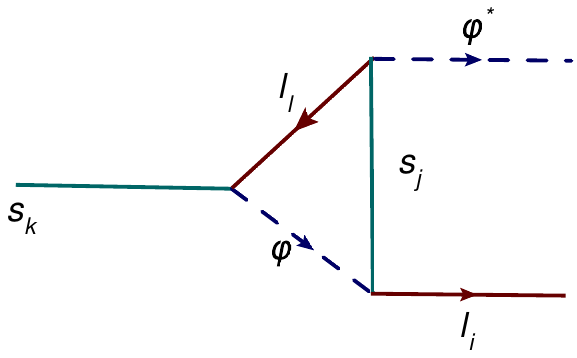}
\includegraphics[width=4 cm,height=3 cm]{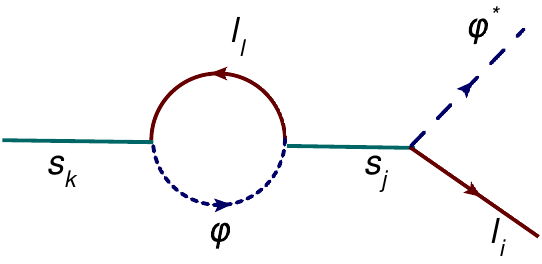}
\caption{Tree and one-loop diagrams for the $S_{k}$ decay contributing to the CP-asymmetry. All fermion-Higgs couplings in the diagrams are of the
form $Vh$ where $h= N-l-\Phi$ Yukawa coupling and $V\simeq M/M_N$.}
\label{fig:Fvertex}
\end{center}
 \end{figure}

The CP-asymmetry formula for the resonant leptogenesis is\cite{bpnmkp:2015}
\bea
\varepsilon_{S_k}&=& \sum_j\frac{{\cal I}m[(y^{\dagger}y)_{kj}^2]}{
  |y^{\dagger}y|_{jj}|y^{\dagger}y|_{kk}}R  \nonumber\\
R&=&\frac{({\hat m}_{S_i}^2-{\hat m}_{S_j}^2){\hat
    m}_{S_i}\Gamma_{S_j}}{({\hat m}_{S_i}^2-{\hat m}_{S_j}^2)^2
+{\hat m}_{S_i}^2\Gamma_{S_j}^2} \,,
\label{epsN}
\eea
where $y= (M/M_{N})h$, $h=M_D/V_{\rm wk}$ ,and $V_{\rm wk}\simeq
174$ GeV.
In order to estimate lepton asymmetry caused by the decay of heavy
sterile fermions ${\hat S}_k (k=1,2,3)$ via their mixing with the heavier RH
neutrinos, the corresponding Feynmann diagrams at the tree and one-loop levels, including the
 vertex and self energy diagrams, are
 shown in Fig. \ref{fig:Fvertex}.

The fermion-Higgs coupling in all the diagrams is
 $Vh$ instead of the standard Higgs-Yukawa coupling 
$h=M_D/ V_{\rm  wk}$ where  ${\mathcal V}\simeq{M/M_N}$,
$M_D$ is given in eq.(\ref{MD}), and $V_{\rm  wk}\simeq 174$ GeV.  
Denoting the mass eigen value of a sterile neutrino by ${\hat m}_{S_k}
(k=1, 2, 3)$,
for computation of  baryon asymmetry $Y_B$ of the Universe with a  washout
factor $K_k$, we utilise the ansatz
\cite{pilaftsis:2003} 
\bea
Y_B &\simeq& \frac{\varepsilon_{S_k}}{200 K_k},\nonumber\\ 
K_k &=&\frac{\Gamma_{S_k}}{H({\hat m}_{S_k})}, \label{bau}  
\eea
$H({\hat m}_{S_k})$ being the Hubble parameter at temperature ${\hat m}_{S_k}$.
Defining 
\be
\delta_i=\frac{|{\hat m}_{S_i}-{\hat m}_{S_j}|}{\Gamma_{S_i}}(i\neq
j), 
\ee
the depleted washout factor is \cite{Hambye}
\be
K_i^{\rm eff}\simeq \delta_i^2 K_i.\label{keff}
\ee 
Here we discuss two cases for the sterile neutrino contribution towards leptogenesis and baryon asymmetry:
(a)  ${\hat m}_{s_1}$ is light, ${\hat m}_{s_2}$ and
${\hat m}_{s_3}$ are quasi-degenerate;
(b) ${\hat m}_{s_2}$ is light, ${\hat m}_{s_1}$ and  ${\hat m}_{s_3}$ are quasi-degenerate.\\
\noindent{\bf{Case (a). ${\hat m}_{s_1}$  light, ${\hat m}_{s_2}$ and  ${\hat m}_{s_3}$ heavy and quasi-degenerate.}}\\
Using an allowed interesting region of the parameter
space $M \simeq {\rm diag.} (146,3500,3500)$ GeV,  eq.(\ref{fQD}),
 eq.(\ref{matms}),  $V_R=10^4$ GeV, and $M_N=fV_R$  we get
\bea 
{\hat m}_{S_i} &=& {\rm diag.} (1.0, 595.864.., 595.864..){\rm
  GeV}.\label{msi1}
\eea 
leading to
 $K_2= 2.7\times 10^7$.  Using $({\hat m}_{S_2}-{\hat m}_{S_3})\simeq
2\times 10^{-7}$ GeV, we obtain
\bea
\varepsilon_{S_2}&=& 0.824,\nonumber\\
Y_B&=& 1.5\times 10^{-10}.\label{bau2f1}
\eea

\par\noindent{\bf{Case (b) ${\hat m}_{s_2}$ light,  ${\hat m}_{s_1}$ and  ${\hat m}_{s_3}$ heavy
    and quasi-degenerate.}}\\
 
Choosing another allowed region of the parameter space $M \simeq {\rm
  diag.} (3200,146,3200)$ GeV, similarly we get
 \bea 
{\hat m}_{S_i} &=& {\rm diag.} (500.567.. ,1.0, 500.567..){\rm
  GeV}.\label{msi2}
\eea
leading to
 $K_1= 4\times 10^6$.  using $({\hat m}_{S_1}-{\hat m}_{S_3})\simeq
7\times 10^{-5}$ GeV, we obtain
\bea
\varepsilon_{S_1}&=& 0.682,\nonumber\\
Y_B&=& 4\times 10^{-10}.\label{bau2f2}
\eea

In Case (a) with  ${\hat m}_{S_1}\sim {\mathcal O}(1)$ GeV , the lightest sterile
 neutrino acts as the most
 dominant source of  $0\nu\beta\beta$ decay whereas the heavy quasi-degenerate
pair of sterile neutrinos $S_2$ and $S_3$ mediate resonant
leptogenesis. Similarly in the alternative scenario  of Case (b) with ${\hat
  m}_{S_2}\sim {\mathcal O}(1)$ GeV, the second generation light
sterile neutrino acts as the  mediator of dominant double beta decay
while the heavy quasi-degenerate pair of the first and the third
generation sterile neutrinos mediate resonant leptogenesis. 
Because of the resonant leptogenesis constraint, we note that either
Case (a) or Case (b) is permitted, but not both.   

\begin{table}
\begin{tabular}{|c|c|c|c|c|}
\hline
$m_{s_1}$&$m_{s_2}$&$m_{s_3}$&Baryon &$T_{1/2}^{0\nu}$\\
 (GeV)&(GeV)&(GeV)&asymmetry&$(10^{25}yrs.)$\\ \hline
 1 & 500 & 500 &$3.73\times10^{-10}$ & 2.72 \\ \hline
10 & 500 & 500 &$3.5\times10^{-10}$ & 16.01 \\ \hline
 500 & 1 & 500 &$4.2\times10^{-10}$ & 0.0494\\ \hline
 500 & 3 & 500 & $4.1\times10^{-10}$ & 2.19\\ \hline
\end{tabular}
\caption{Predictions for baryon asymmetry and double-beta decay
  half-life as a function of sterile neutrino masses.}
\label{bsymm}
\end{table}
 Our predictions for the double beta decay half-life and the baryon
 asymmetry in Case (a) and Case (b) are presented in Table
 .\ref{bsymm}. It is clear that for smaller mass eigen values of
 sterile neutrinos in Case (a) or Case (b), it is possible to saturate
 current experimental limit on the double-beta decay half-life while
 explaining the right order of magnitude of the baryon
 asymmetry. Thus, in addition to the Case (a) found in
 ref.\cite{bpnmkp:2015}, we have shown another possible alternative scenario as
 Case (b).

\section{\bf DECAY WIDTH, HALF-LIFE, AND  DISPLACED LENGTHS OF STERILE
  NEUTRINOS}
\begin{figure}[htbp]
 \includegraphics[width=6cm,height=4cm]{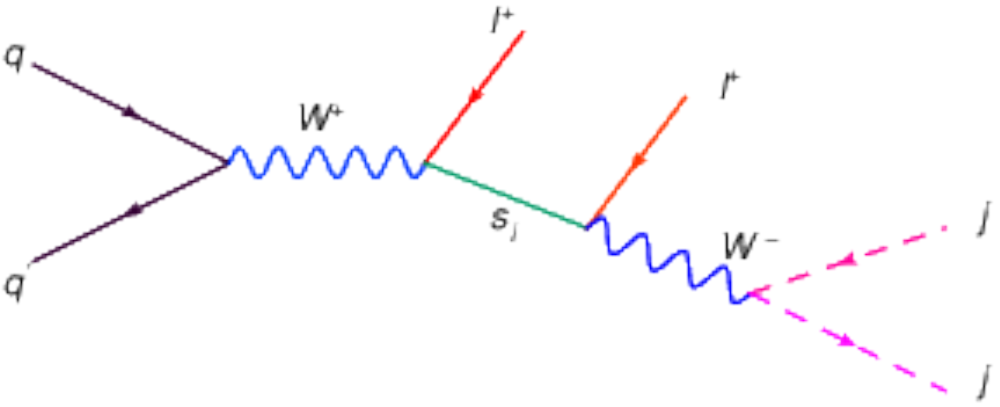}
 \caption{ The Feynman diagram for like-sign dilepton production process
$l^{\pm}l^{\pm}jj$ in the $W_L-W_L$ channel in the $pp$ collision process at LHC}
\label{feyn}
 \end{figure}

When the RH Majorana neutrino is near the TeV scale with shorter lifetime and
path length, it is expected to mediate like-sign dilepton events
inside the LHC detectors \cite{keung,saavedra:2009,cdm:2013}. But 
when the mass of the Majorana type sterile neutrino is ${\cal O}(1)$
GeV having its 
 naturally small mixing with
active light neutrinos, its half life is longer 
 resulting in its path lengths extending to regions
outside the LHC detectors. This is expected to cause the dilepton signal events
to be observed via displaced vertices outside the detectors. As already explained, because of the
$SO(10)$ model predictions of heavy-light mixing matrix elements
  ${\cal V}_{ei},{\cal V}_{\mu i} ~(i=1,2)$, either $S_1$ or $S_2$ is capable of mediating the displaced vertices resulting in the like-sign dilepton events $eejj$, $\mu\mu jj$, and $e\mu jj$ without missing energy and with almost negligible 
SM background provided that the signal strength is strong enough under different cut conditions \cite{Helo}.
 The Feynman
 diagram for the production of like-sign dileptons along with two jets in the $W_L-W_L$ channel is given in Fig.(\ref{feyn}). The 
contribution of sterile neutrino to neutrino-less double beta decay is
shown in Fig.(\ref{feyn1}). As already clarified the double beta decay
occurs even with the exchange of ${\hat S}_2$ because of the existence
of the elements ${M_D}_{e2}$ in eq.(\ref{MD}) that gives rise to quite
significant mixing ${\cal V}_{e2}$. 
More specifically, because of the 
possibility of an alternative case shown as Case (b) and
Table. \ref{bsymm} with lighter value of ${\hat m}_{S_2} \sim
{\cal O}(10)$ GeV and two other heavy and quasi-degenerate mass eigen
values, it would be interesting to explore the outcome of the Case (b)
along with the Case (b) for dilepton production with displaced vertices.
\\

 At LHC the
sterile neutrino mediated cross section can be 
expressed in terms of heavy-light mixing \cite{pascoli} \\

 $\sigma(pp\rightarrow Sl^{\pm} \rightarrow l^{\pm} l^{\pm} jj)=(2-\delta_{l1l2})S_{l1l2}\sigma_{0} (S)$\\

 \bea
 S_{l_1l_2}=\frac{|V_{l_{1}s} V_{l_2}s|^2}{\sum_{l=e,\tau}|V_{ls}|^2}
 \eea
 where $l=e~{\rm or}~\mu$  and $\sigma_{0}(S)$ is the bare cross
 section arising out of the exchange of the sterile neutrino $S$.

If the second lepton is produced outside the LHC detector within a path length
$L$ defined by eq.(\ref{length}) given below,  
 the number of events within the displaced length limit is estimated through 
the formula \cite{Helo}
 \bea
 N = L \times \sigma(pp\rightarrow Sl^\pm \rightarrow l^\pm jj) \times PN \\
 PN = e^{-{d_{1}/L}}-e^{-{d_{2}/L}}
 \eea

In the Type-I seesaw based RH sterile neutrino case \cite{Helo}, using the
DELPHI bound \cite{DELPHI} and double beta constraint on LH and RH 
 neutrino mixing, a
correlation between sterile neutrino mass and allowed values of
mixings has been derived under different cut conditions \cite{ATLAS} such that
approximately $3-5$ like-sign dilepton events in the $l^{\pm}l^{\pm}jj$ channel
can  be detected outside the LHC detectors with luminosity ${\cal L}
= 300$ fb$^{-1}$. Keeping the transverse momentum cut of the
first electron at $p_T^{e_1} > 30 $ GeV, the allowed region sensitive to mixings has
been identified under the momentum cut conditions for the second
electron for $p_T^{e_2} > 7, 30, 35, ~{\rm and},~45$ GeV with rapidity
cut  $|\eta^{e_2}| < 2.5$. Also keeping  $p_T^{e_2} > 7$ GeV, the
allowed region has been investigated under the jet momentum cut
conditions  $p_T^j > 10, 15, ~{\rm and}~ 20 $ GeV for  $p_T^{e_1} > 30 $
GeV and $|\eta^{e,j}| < 2.5$. It turns out that  $3-5$ events of
$eejj$ might be barely possible in the double beta decay mixing region 
with $|V_{ls_i}|^2\simeq 10^{-7}-10^{-8}$ provided a lower value of 
the  second leptonic
momentum cut is imposed and the luminosity is large enough, ${\cal L}
= 3000$ fb$^{-1}$. However for larger values of mixings displaced
vertices can be observed for lower luminosity like   ${\cal L}
= 300$ fb$^{-1}$.  For the RH sterile neutrino mass
in the range $2-80$ GeV,
  subject to different cut conditions, the  assumed values of modulus square of
mixings sensitive for displaced vertex search have been found to be 
constrained to the region  $10^{-5}$ to $10^{-7}$ .  

In our Model-I and Model-II, the heavy light mixings as well as the
sterile neutrino masses are predicted by the underlying mechanism in $SO(10)$. We investigate how these model predictions are accommodated in the almost model-independent approach
of ref. \cite{Helo}. We also predict the number of events that can be
produced through  displaced vertices in other channels like $e\mu jj$,
$\mu\mu jj$, and $e\tau jj$.    
The decay width of the i-th light sterile neutrino $(i=1~{\rm or}~2)$  of these  two models in Case (a) or Case (b) as discussed above is
\bea 
\Gamma_{s_i}=\frac{3G_{F}^2}{32\pi^{3}} m_{s_i}^5\sum_{l} |{\cal V}_{ls_i}|^2,
\eea
where $G_F=$  Fermi coupling constant and ${\cal V}_{l{s_i}}$ is the 
light neutrino-
sterile neutrino mixing matrix.\\
Coexisting with sterile neutrinos $S_1, S_2, S_3$ in Model-I and Model-II,
the heavy RH neutrinos of three generations with masses at the $\sim $
TeV scale are assumed to predict
negligible new contributions to double beta decay and also to
 displaced vertices outside the LHC detectors, 
ATLAS or CMS. In contrast, one of the sterile neutrinos being
sufficiently light, $S_1$ in Case (a) or $S_2$ in Case (b) in each of the two
models (Model-I and Model-II),  
is expected to provide quite effective mediation
for these two processes.
With the value $G_{01}=0.686\times 10^{-14} {\rm  yr}^{-1}$, the
nuclear matrix element ${\cal
  M}_N=233-412$, $m_p=$ proton mass
  defined
through 
eq.(\ref{invhalf}), the half-life of the i-th  sterile neutrino of mass eigen value
${\hat m}_{s_i}$ is\\
\bea
T_{1/2}^{-1}=G_{01}{({\cal M}_N m_p)}^2 {{\cal V}_{ls_i}}^4{\hat m}_{s_i}^{-2}
\eea
leading to the displaced length of the first generation sterile
neutrino  $S_1$ in case (a) or  the second generation sterile neutrino
$S_2$ in Case (b)\\
\bea
L=4875 {\bar{\gamma}}_i (\frac{GeV}{{\hat m}_{s_i}})^5  \frac{10^{-7}}{|V_{ls_i}|^2}
, \label{length}
\eea
where ${\bar{\gamma}}_i= \frac{E_i}{{\hat m}_{s_i}}$, 
$E_i$ being the average energy of the sterile neutrino $S_i$.\\

 From  DELPHI \cite{DELPHI} the mixing limit is
$\sim 10^{-5}$  and from the neutrino-less double beta
 $(0\nu\beta\beta)$ decay the upper limit (lower limit) of mixing is
 $10^{-7} (10^{-8})$.
Our model is different from other models in that the heavy-light
mixings, be it with RH neutrinos or sterile neutrinos, are determined
from the fermion mass fits at the $SO(10)$ GUT scale and the LFV
constraint on $M_i (i=1,2,3)$ \cite{bpnmkp:2015}.
  As already noted, the Dirac neutrino mass
matrix in eq.(\ref{MD}) that is basic to heavy-light mixing has been determined
 by fitting the RG extrapolated values of charged fermion
masses at the GUT scale and running the matrix elements to the TeV
scale using the top-down approach. 
 Then the $\nu-N$ mixing matrix is $\frac{M_D}{M_N}$ whereas the
$\nu-S$ mixing matrix is $\frac{M_D}{M}$.\\  

\begin{figure}[htbp]
\includegraphics[width=7cm,height=5cm]{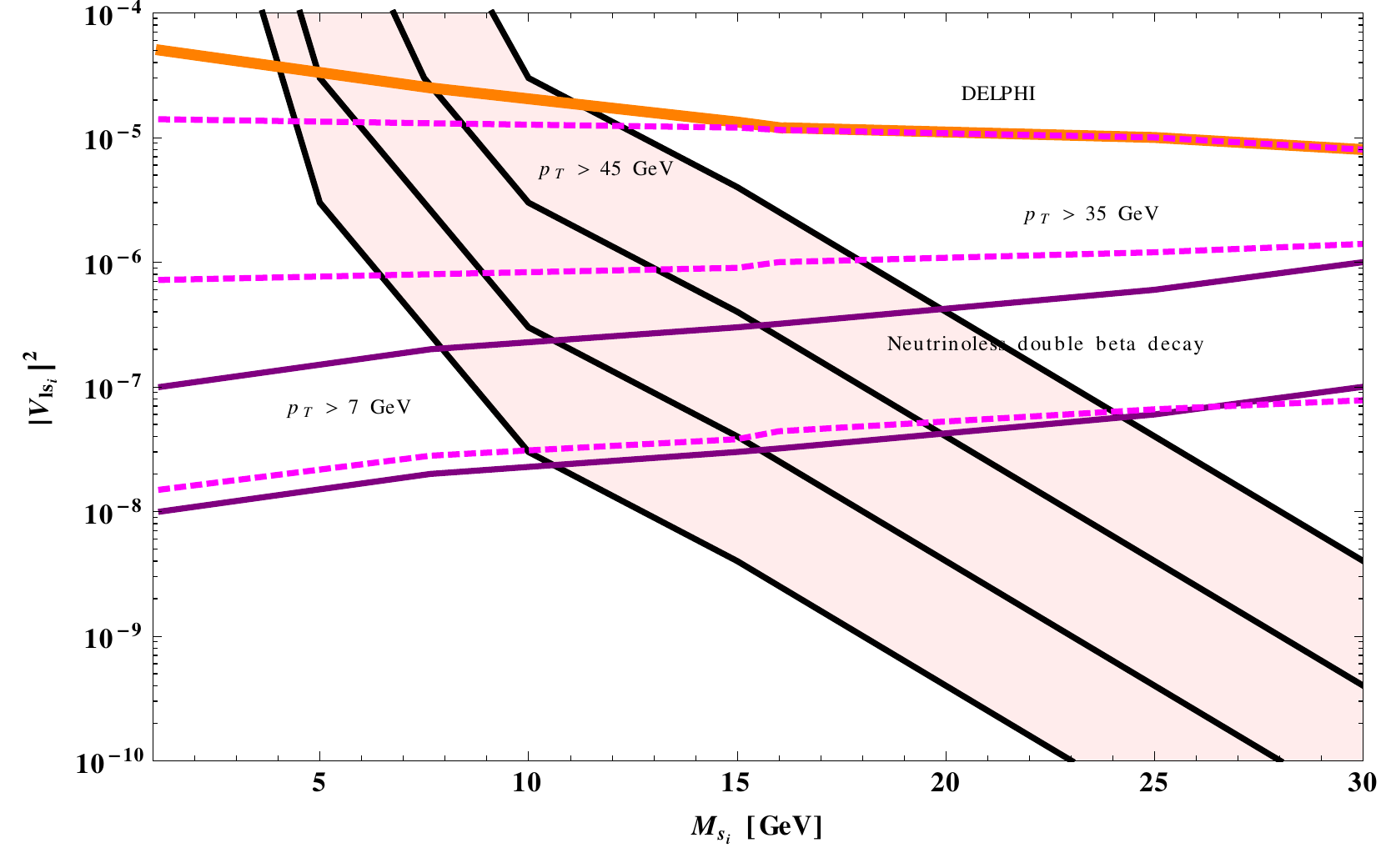}
\caption{Constraints on active-sterile neutrino mixings including DELPHI and neutrinoless double beta decay limits. The lower-most solid line of the pink colored shadow region having upward concavity corresponds to L=1m and the upper-most line corresponds to $L=0.001$ m where the displaced vertex search at LHC is 
expected to be sensitive.
The dashed lines correspond to different $p_T$ cuts: $p_T^{e_1}> 40$ GeV, $p_T^{e_2}> 7$ GeV, $p_T^{e_1}> 40$ GeV, $p_T^{e_1}>45$ GeV
and $|\eta^{e}| < 2.5$ with luminosity $300 fb^{-1}$. The two lower solid curves
with positive slopes indicate the double beta decay limits.}
\label{pdrn1}
\end{figure}

The model-independent analysis of the type given in ref.\cite{Helo}
using the formula of eq.( \ref{length}) is shown in  Fig.\ref{pdrn1}.
The mass of the sterile neutrino in the range $1-80$ GeV, 
corresponds to the displaced length
of the order of $(0.001-1)$m and the displaced vertex search is sensitive in the pink coloured shadow region 
given of Fig.\ref{pdrn1}. \footnote{Compared to the corresponding straight line
curve for a fixed displaced length $L$ shown in
Fig.1 of \cite{Helo}, each of our four curves show 
a bending with upward concavity at $M_{S_i} \sim 10$ GeV. This difference arises because of choice of two different scales in \cite{Helo} compared to one uniform scale for the mass axis in this work.} 
 
An important point shown in this work is that the heavy-light mixings predicted
by our Mode-I and Model-II based on $SO(10)$, either in Case (a) or Case (b) of each model, 
are found to be in the sensitive region of the model-independent
search \cite{Helo}. In the following section we predict signal events
of dilepton production due to displaced vertices in various channels.

\vspace {4cm}
 \section{ MODEL PREDICTION OF DILEPTON SIGNAL EVENTS WITH DISPLACED VERTICES}
In this section we estimate the number of signal events that may
appear outside the LHC detectors using the
heavy-light neutrino mixings and the sterile neutrino masses predicted
in our Model-I and Model-II in Case (a) and Case (b).   
 In earlier ananlyses with the single  RH neutrino as sterile
 neutrino, like-sign dilepton events have been investigated mostly in
 the $eejj$ channel \cite{Helo} for various assumed values of mixings
 satisfying experimental constraints from  DELPHI and double-beta
 decay experimental limits. In the present work the, because of the available predictions
 on two interesting
 possible cases for double beta decay and resonant leptogenesis
 caused by the lighter masses of the  sterile neutrino $S_1$ or $S_2$,
 we are able to predict like-sign dilepton events for $eejj$, $\mu\mu
 jj$, $e\mu jj$ , and $e\tau jj$ channels as shown in  Table. \ref{tab1:die},
Table. \ref{tab2:dimu},  Table. \ref{tab3:dimu},  Table. \ref{tab4:die},
Table. \ref{tab5:diemu},   Table. \ref{tab6:diemu}, and
Table. \ref{tab7:diet} for the respective cases as functions of
sterile neutrino masses, luminosities, and cut conditions.
As clarified through our analytic formulas in eq.(\ref{matms}) and 
eq.(\ref{HLMIX}), the sterile neutrino mass eigen value can be
increased or decreased by keeping $M_1$ or $M_2$ fixed over limited
range of values while decreasing or increasing $M_N$ that depends upon
the ratio of two VEVs, ${V_R\over v_L}$, Since our sterile
neutrino-light neutrino mixings are inversely proportional  to $M_i$,
this ratio has been also utilised to have the desired values of
mixings for larger values of sterile neutrino masses.

\begin{table}[htbp]
\begin{tabular}{|c|c|c|c|c|}
\hline
Cut&${\hat m}_{s_1}$&${V_R/v_L}$&$L(300)$&$L(3000)$\\
Condition&(GeV)& & $(fb^{-1})$& $(fb^{-1})$\\ \hline
$P_T^{e_2}\geq 7$ GeV & & & &\\
 $P_T^{e_1}\geq 30$ GeV &1.2 & $2\times10^{13}$& 0.5& 5\\ \hline
$P_T^{e_2}\geq 7$ GeV& & & &\\
 $P_T^{e_1}\geq 30$ GeV & 6 & $10^{14}$& 0.5& 5\\ \hline
\end{tabular}
\caption{Prediction of like sign dielectron events via displaced vertices in the $eejj$ channel as a function of sterile neutrino mass ${\hat m}_{s_1}$.}   
\label{tab1:die}
\end{table}
\begin{table}[htbp]
\begin{tabular}{|c|c|c|c|c|}
\hline
Cut&${\hat m}_{s_2}$&${V_R/v_L}$&$L(50)$&$L(300)$\\
Condition&(GeV)&& $(fb^{-1})$& $(fb^{-1})$\\ \hline
$P_T^{\mu_2}\geq 35$ GeV&& & &\\
 $P_T^{\mu_1}\geq 30$ GeV &1.2 &$2\times10^{13}$& 2.67& 16\\ \hline
$P_T^{\mu_2}\geq 35$ GeV& & &&\\
 $P_T^{\mu_1}\geq 30$ GeV &6 &$10^{14}$& 3& 16\\ \hline
\end{tabular}
\caption{Prediction of like sign dimuon events via displaced vertices in the $\mu\mu jj$ channel as a function of sterile neutrino mass ${\hat m}_{s_2}$.}   
\label{tab2:dimu}
\end{table}
\begin{table}[htbp]
\begin{tabular}{|c|c|c|c|c|}
\hline
Cut&${\hat m}_{s_1}$&${V_R/v_L}$&$L(50)$&$L(300)$\\
Condition&(GeV)&&$(fb^{-1})$& $(fb^{-1})$\\ \hline 
$P_T^{\mu_2}\geq 35$ GeV&& & &\\
 $P_T^{\mu_1}\geq 30$ GeV &1.2 &$2\times10^{13}$ &2& 9\\ \hline
$P_T^{\mu_2}\geq 35$ GeV& &&&\\
 $P_T^{\mu_1}\geq 30$ GeV &6 &$10^{14}$& 2& 9\\ \hline
\end{tabular}
\caption{Prediction of like sign dimuon events via displaced vertices in the $\mu\mu jj$ channel as a function of sterile neutrino mass ${\hat m}_{s_1}$.}   
\label{tab3:dimu}
\end{table}
\begin{table}[htbp]
\begin{tabular}{|c|c|c|c|c|}
\hline
Cut&${\hat m}_{s_2}$&${V_R/v_L}$&$L(300)$&$L(3000)$\\
Condition&(GeV)&&$(fb^{-1})$& $(fb^{-1})$\\ \hline 
$P_T^{e_2}\geq 35$ GeV&& & &\\
 $P_T^{e_1 }\geq 30$ GeV &1.2 &$2\times10^{13}$&0.51&5.1\\ \hline
$P_T^{e_2}\geq 35$ GeV& & & &\\
$P_T^{e_1}\geq 30$ GeV &6 &$10^{14}$ &0.51&5.1\\ \hline
\end{tabular}
\caption{Prediction of like sign dielectron events via displaced vertices in the $eejj$ channel as a function of sterile neutrino mass ${\hat m}_{s_2}$.}   
\label{tab4:die}
\end{table}
\begin{table}[htbp]
\begin{tabular}{|c|c|c|c|c|}
\hline
Cut&${\hat m}_{s_1}$&${V_R/v_L}$&$L(300)$&$L(3000)$\\
Condition&(GeV)&&$(fb^{-1})$& $(fb^{-1})$\\ \hline 
$P_T^{e_2}\geq 35$ GeV&& & &\\
 $P_T^{e_1 }\geq 30$ GeV &1.2 &$2\times10^{13}$&0.3 &3.26\\\hline
$P_T^{e_2}\geq 35$ GeV& & & &\\
$P_T^{e_1}\geq 30$ GeV &6 &$10^{14}$ &0.3 &3.26\\\hline
\end{tabular}
\caption{Prediction of electron-muon events via displaced vertices in the $e\mu jj$ channel as a function of sterile neutrino mass ${\hat m}_{s_1}$.}   
\label{tab5:diemu}
\end{table}
\begin{table}[htbp]
\begin{tabular}{|c|c|c|c|c|}
\hline
Cut&${\hat m}_{s_1}$&${V_R/v_L}$&$L(300)$&$L(3000)$\\
Condition&(GeV)&&$(fb^{-1})$& $(fb^{-1})$\\ \hline 
$P_T^{e_2}\geq 35$ GeV&& & &\\
 $P_T^{e_1 }\geq 30$ GeV &1.2 &$2\times10^{13}$&0.58& 6\\\hline
$P_T^{e_2}\geq 35$ GeV& & & &\\
$P_T^{e_1}\geq 30$ GeV &6 &$10^{14}$ &0.5&6\\\hline
\end{tabular}
\caption{Prediction of electron-muon events via displaced vertices in the $e\mu jj$ channel as a function of sterile neutrino mass ${\hat m}_{s_2}$.}   
\label{tab6:diemu}
\end{table}
\begin{table}[htbp]
\begin{tabular}{|c|c|c|c|c|}
\hline
Cut&${\hat m}_{s_2}$&${V_R/v_L}$&$L(50)$&$L(100)$\\
Condition&(GeV)&&$(fb^{-1})$& $(fb^{-1})$\\ \hline 
$P_T^{e_2}\geq 35$ GeV&& & &\\
 $P_T^{e_1 }\geq 30$ GeV &1.2 &$2\times10^{13}$&28  &56\\\hline
$P_T^{e_2}\geq 35$ GeV& & & &\\
$P_T^{e_1}\geq 30$ GeV &6 &$10^{14}$&28 &56\\\hline
\end{tabular}
\caption{Prediction of electron-taon events via displaced vertices in the $e\tau jj $ channel as a function of sterile neutrino mass ${\hat m}_{s_2}$.}   
\label{tab7:diet}
\end{table}

We conclude that dilepton events through displaced vertices are
possible in various channels with suppressed SM back ground events and
also with vanishing missing energy. This serves as an interesting
prediction based upon Type-II seesaw dominant $SO(10)$ GUTs.  

\section{DILEPTON SIGNATURE BY HEAVY RH NEUTRINO EXCHANGE}
After discussing the manifestation of sterile neutrinos in Model-I and
Model-II through various physical processes like double beta decay,
dilepton signals with displaced vertices, and resonant leptogenesis,
in this section we investigate if the TeV scale RH neutrinos present in
both the models may manifest at the LHC, particularly, through the
like-sign dilepton production events that may materialise inside the
ATLAS or the CMS detectors. Since the $W_R$ boson mass is quite heavy 
$M_{W_R} > 10^8$ GeV, only the $W_L-W_L$ channel is dominant for the
process $pp\to l^{\pm}l^{\pm} X$ where $l=e,\mu$.
The heavy RH Majorana neutrino exchange cross section is given by 
\cite{cdm:2013}\\

$\sigma(pp\rightarrow Nl^\pm \rightarrow l^\pm jj)=\sigma_{prod}(pp\rightarrow W_L\rightarrow Nl^\pm) \\
\times BR(N\rightarrow l^\pm jj)$\\
where the production cross section $\sigma_{prod}$ is estimated by using patron 
level distribution function CTEQ6L \cite{pumplin}.
The branching ratio is estimated using \\
\bea
BR(N\rightarrow l^\pm jj)=\frac {\Gamma (N\rightarrow l^\pm jj)}{\Gamma^{tot}_{N}}\times BR(W \rightarrow jj)
\eea
with $BR(W \rightarrow jj)=0.676$. The total width is calculated by sum of all the partial widths\\
\bea
\Gamma(N\rightarrow l^{\pm} W) &=& \frac{g^2|V_{\nu N}|^2} {64
  \pi}\frac{{M_{N}}^3}{{M_W}^2}(1-\frac{{M_{W}}^2}{{M_{N}}^2})^2
\nonumber\\
&&\times(1+\frac{2{M_{W}}^2}{{M_{N}}^2}),
\eea
\bea
\Gamma(N\rightarrow \nu_{l} Z) &=& \frac{g^2|V_{\nu N}|^2} {128 \pi
  Cos^{2}\theta_w}\frac{{M_{N}}^3}{{M_Z}^2}(1-\frac{{M_{Z}}^2}{{M_{N}}^2})^2
\nonumber\\
&&\times (1+\frac{2{M_{Z}}^2}{{M_{N}}^2}),
\eea
\bea
\Gamma(N\rightarrow \nu_{l} h)=\frac{g^2|V_{\nu N}|^2} {128 \pi}\frac{{M_{N}}^3}{{M_W}^2}(1-\frac{{M_{h}}^2}{{M_{N}}^2})^2
\eea
\begin{figure}[htbp]
 \includegraphics[width=8cm,height=6cm]{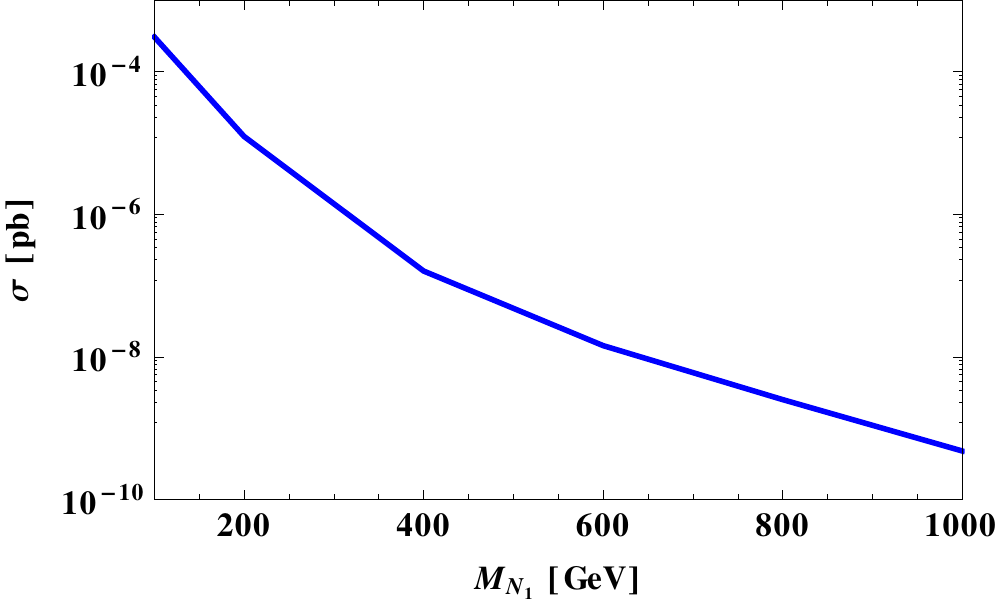}
 \caption{The signal cross section of the heavy RH neutrino as a function of its mass for dimuon at LHC with
 $\sqrt{S}=14$ TeV} .    
\label{dlprhn}
\end{figure}

The heavy-light neutrino mixing plays a  crucial role in calculating the signal cross section. In our model 
the heavy-light neutrino mixing matrix  for heavy RH neutrinos is $\frac{M_D}{M_N}$, where $M_D$ is the Dirac neutrino mass matrix and $M_N$ is the
 RH neutrino mass matrix. An interesting aspect of the models is that
 both the matrices, $M_D$ and $M_N$, are already predicted by TeV
 scale gauge symmetry breaking, the neutrino oscillation data, and
 charged fermion mass fits at the GUT scale. 

In the case of  vanilla seesaw model investigated in \cite{cdm:2013}, the
heavy-light neutrino mixing is $V_{lN}^2=m_{\nu}/M_N $ which is quite
small and gives rise to a  cross section $\sigma(pp \to \mu\mu jj X)
\simeq 10^{-16}$ pb in the  $W_L-W_L$ channel for exchanged heavy RH
neutrino mass $\simeq 100$ GeV.
On the other hand if the heavy-light mixing is assumed to be as large as
$V_{lN}^2=3\times10^{-3}$ \cite{cdm:2013}, the cross section is also large leading to
$\sigma(pp \to \mu\mu jj X) \sim 6\times10^{-2}$ pb. 
However we do not assume any such large mixings here.
In our type-II seesaw dominant models where all the heavy-light
neutrino mixings are predicted, the estimated value of dimuon signal
cross section in the $W_L-W_L$ channel turns out to be   $\simeq 5\times 10^{-4} (2\times 10^{-5})$ pb for the mass 
of $M_{N_1}= 100 (200)$ GeV  
resulting in nearly $24(12)$ events for beam luminosity ${\cal
  L}=300$fb$^{-1}$ after including the cuts\cite{pascoli}.  This result is shown in  fig.(\ref{dlprhn}). Thus if
the RH neutrino masses are within $M_N\le 300$ GeV, they are
detectable at LHC run-II at $\sqrt s = 14$ TeV for projected  beam luminosity ${\cal
  L}=3000$ fb$^{-1}$, although  the RH neutrino masses $M_{N_1}\le 200$ GeV
are detectable with beam luminosity  ${\cal
  L}=300$ fb$^{-1}$. In these models the larger values of RH neutrino masses, 
$M_{N_1} > 500 $ GeV, are likely to escape detection at LHC through
like-sign dilepton production signals. These conclusions remain valid
after imposing the various cut conditions applicable to the $pp\to
l^{\pm}l^{\pm} jj X$ channels \cite{pascoli,cdm:2013}. In the case of $M_{N_2}$
we have similar conclusion. 
\section{\bf DOUBLE BETA DECAY, DISPLACED VERTICES, AND LEPTOGENESIS
  WITH NH NEUTRINOS}
 
\begin{figure}[htbp]
 \includegraphics[width=7cm,height=8cm,angle=0]{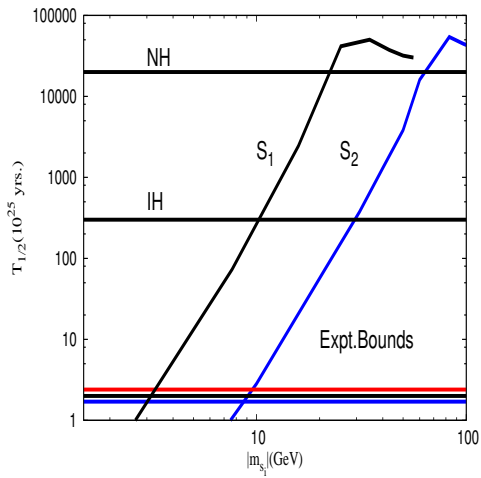}
 \caption{The half-life of the neutrinoless double beta decay as a function of sterile neutrino mass. }   
\label{pdrn}
\end{figure}

In ref.\cite{bpnmkp:2015}, we have already shown how dominant double
beta decay rate occurs for  normally hierarchical (NH), invertedly
hierarchical (IH), and quasi-degenerate (QD) mass patterns  of light and active
neutrinos due to the exchange of lightest sterile neutrino $S_1$. But
our applications as given above for dominant double beta decay
along with
displaced vertices and resonant leptogenesis have been confined only to the 
 QD pattern of light neutrino masses.  On the otherhand if the recent cosmological limit \cite{cosnu:2014} is
 ultimately confirmed by laboratory experiments, neutrino masses could
 be NH type. It would be interesting to investigate if the results
 derived so far are also applicable to NH case.
For this purpose we note that in the NH case the new contribution to
$0\nu\beta\beta$ half-life by $S_1$ saturates the experimental bound
for ${\hat m}_{S_1}= 2.5\pm 0.5$ GeV as shown  in
fig.(\ref{pdrn}). Such a mass certainly mediates like-sign dilepton
events with displaced vertices in the $eejj$ and $\mu\mu jj$ channels
as discussed above. To complete the application to resonant
leptogenesis we search for parameter values in $M={\rm diag.}
(M_1, M_2, M_3)$ to find solutions for quasi-degenerate pair of masses ${\hat m}_{S_2}$ and  
${\hat m}_{S_3}$ near TeV scale so that the ansatz given above and  in ref.
\cite{bpnmkp:2015} goes through to explain baryon asymmetry of the universe.

 We choose an interesting region of the parameter
space $M \simeq {\rm diag.} (38.27, 752.1, 1219.0)$ GeV.
 Then using the RH neutrino mass matrix $M_N$ derived in
 ref.\cite{bpnmkp:2015} in the NH case and  eq.(\ref{matms}), we get
\bea 
{\hat m}_{S_i} &=& {\rm diag.} (2.8, 1348.86.., 1348.86...){\rm
  GeV}.\label{msi}
\eea 
containing the desired mass patterns.
Using the formula for resonant  leptogenesis, we get
 $K_1= 2\times 10^{12}$.  using $({\hat m}_{S_2}-{\hat m}_{S_3})\simeq
0.001$ GeV, we obtain
\bea
\varepsilon_{S_2}&=& 0.013,\nonumber\\
Y_B&=& 7.4\times 10^{-10}.\label{bau2f}
\eea
 The values of ${\hat m}_{S_1}$ upto ${\cal O}(10)$ GeV are easily
 obtained while satisfying the required constraints for resonant leptogenesis.
 With the value of $S_1$ mass given in eq.( \ref{msi}) we have
 verified that the double beta decay lifetime is predicted with a
 value close  to the
 current experimental limit. 

Thus we have shown that for NH pattern of light neutrino masses that
favours the recent cosmological bound \cite{cosnu:2014}, a light
sterile neutrino mass $\sim 2.8$ GeV allowed in both the $SO(10)$ models
mediates dominant double beta decay and displaced vertices for
like-sign dilepton production in the channels $eejj$ and $\mu\mu jj$
In this case resonant leptogenesis is implemented by a
  pair of quasi-degenerate heavy masses $\simeq 1348.86...$ GeV of the
  sterile neutrinos of the second and the third
  generations which is also permitted within the allowed parameter space.

\section{ $Z^{\prime}$ DETECTION AT COLLIDERS}
One important and interesting feature of this paper is the
prediction of of extra
neutral $Z'$  boson  at the TeV scale accessible for
detection at LHC, International Linear
Collider [ILC], and  future collider experiments. In this section we
estimate relevant cross sections which may help in identifying the
$Z'$-boson signals.  

 \subsection{Cross section of $Z^{\prime}$ boson}
In this section we discuss the possible signatures of the $Z^{\prime}$
boson at LHC and ILC experiment through opposite sign
dilepton production cross sections. We also discuss possible signature of
$Z^{\prime}$ boson through the production of $W^+W^-$ pairs for
different values of $Z-Z^{\prime}$ mixings at LHC. In the dilepton
channel we compare our estimated
$Z^{\prime}$ production cross section with those obtained by 
 CMS experiment \cite{CMS:zprime} in the channel
 $pp\to Z'X\to l^{+}l^{-}X$ where $l=e,\mu$.

 The resonant production cross section for the opposite sign dilepton
 production via  $Z^{\prime}$ boson resonance is \cite{carena,Accomando}\\
 \bea
 \sigma(pp\rightarrow Z^{\prime}\rightarrow f\bar f)=\frac{\pi}{48S}[C_{u} w_{u}(S,M_{Z^{\prime}}^2)\nonumber\\+C_{d} w_{d}(S,M_{Z^{\prime}}^2)],
 \eea
 where the coefficients $C_{u}$ and $C_{d}$ are  
 \bea
 c_{{u},{d}}=g_{z}^2(z_{q}^2+z_{{u},{d}}^2)Br(l^{+}l^{-}).
 \eea
Here $w_{{u}{(d)}}$  \cite{Accomando} is related to the parton luminosities $\frac{dL_{u\bar{u}}}{dM_{z'^2}}$ and
  $\frac{dL_{d\bar{d}}}{dM_{z'^2}}$. Therefore  they depend only upon
the collider energy and the $Z^{\prime}$ mass.

\begin{figure}[htbp]
\includegraphics[width=7cm,height=5cm,angle=0]{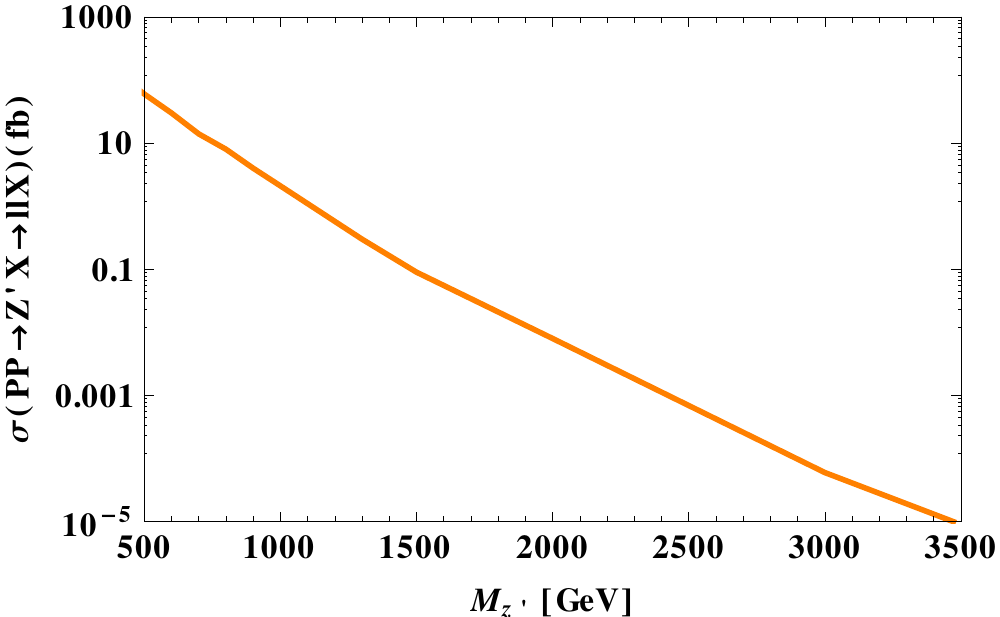}
\caption{The signal cross section of the $Z^{\prime}$ boson as a
  function of its mass in the production channel $pp\to Z'X \to
  l^{\pm}l^{\mp} X$ }
\label{Fig.zpprod}
\end{figure}

The production cross section in the channel $pp \to Z' X \to
l^{\pm}l^{\mp} X$  as a function of invariant dilepton mass (or the $Z'-$ mass) is shown in
Fig.\ref{Fig.zpprod}. This result  suggests that at $\sqrt S= 14$ TeV
the number of $Z'$ production events could be large in the region of
$M_{Z'}\sim 1$ TeV  even for $30$fb$^{-1}$ beam luminousity, but to get sizeable number of events in the region
of $M_{Z'}\sim 2-3$ TeV, the beam luminousity has to increase beyond several
$1000$ fb$^{-1}$.\\


\begin{figure}[htbp]
\includegraphics[width=7cm,height=5cm,angle=0]{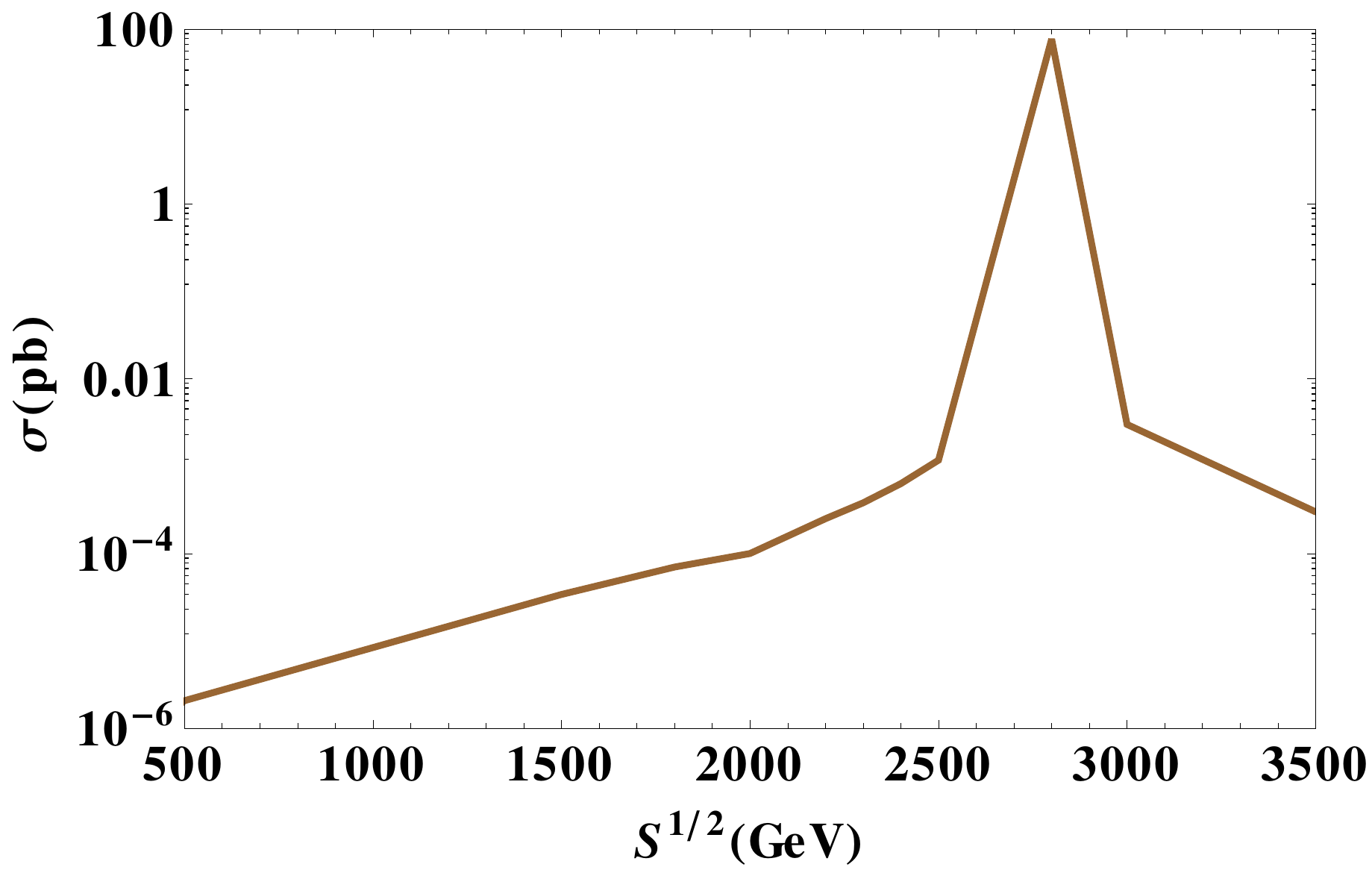}
\caption{The signal cross section of the $Z^{\prime}$ boson as a
 function of its center of mass energy}
\label{zpilc}
\end{figure}

The  international linear collider (ILC) is expected to provide a
rigorous experimental verification of various  $Z^{\prime}$ models as
far as their predicted masses are concerned.  In our model, the
variation of the predicted annihilation cross section via $Z'$
resonance  with center of mass energy of the colliding lepton beams is
given in  Fig.\ref{zpilc}. To estimate this cross section we have used
the total decay width of $Z^{\prime}$ boson as the sum of decay 
widths of $Z^{\prime}$ into quarks and leptons 
\cite{Accomando}. 

\bea
\Gamma_{Z^{\prime}}=\frac{g^{\prime 2}}{48\pi}[9(g_V^{u^2}+g_A^{u^2})+9(g_V^{d^2}+g_A^{d^2})\nonumber\\+3(g_V^{e^2}+g_A^{e^2})+3(g_V^{\nu^2}+g_A^{\nu^2})]
\eea

The signal cross section is found to be $70$ pb at
the center of mass energy $\sqrt s=2800$ GeV which corresponds to the
$Z'$ mass.  Therefore its presence can be easily detected even for as
low a luminousity as ${\cal O}(1)$ pb$^{-1}$.

\begin{figure}[htbp]
\includegraphics[width=9cm,height=7cm,angle=0]{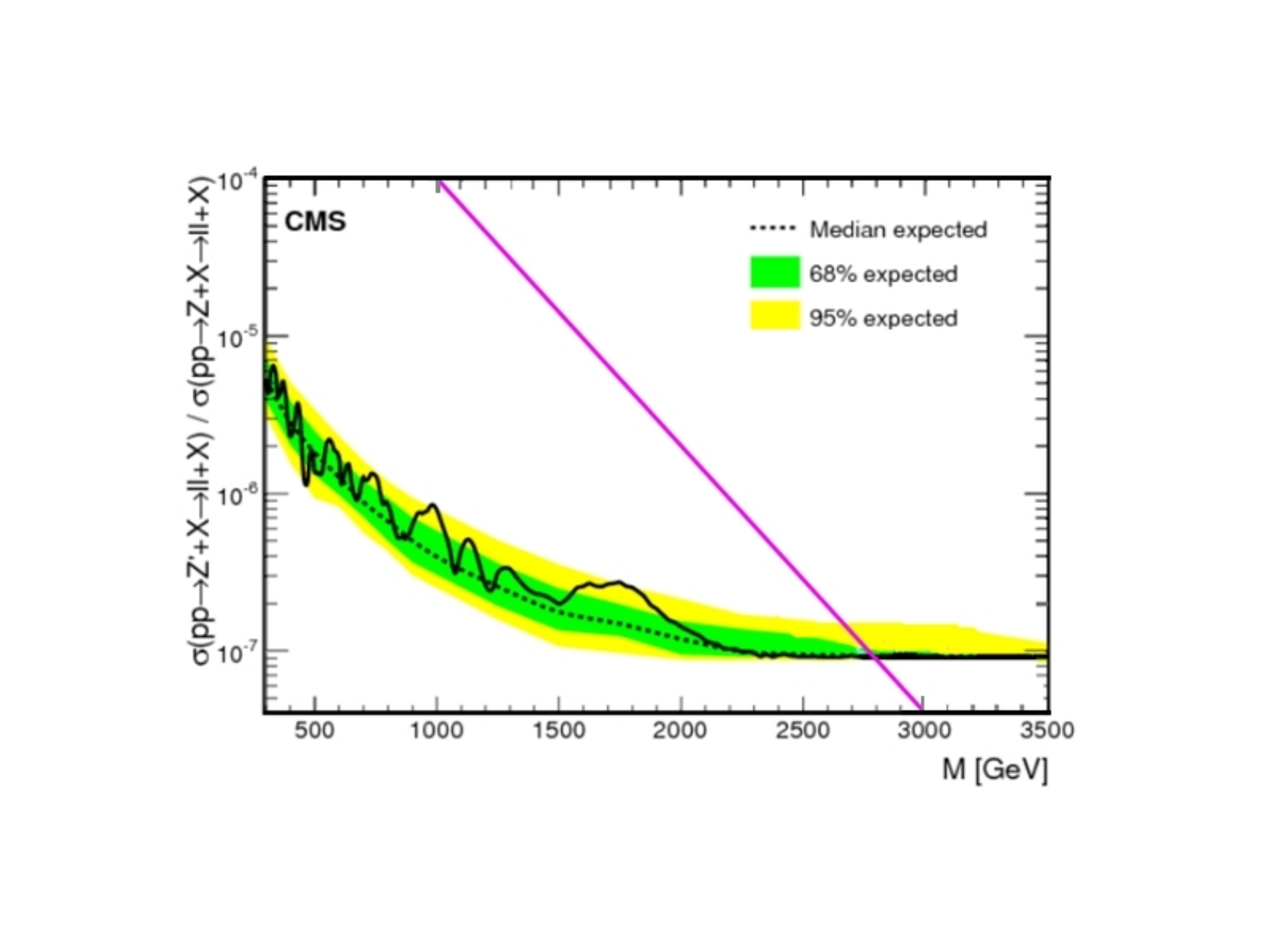}
\caption{The ratio of  signal cross section of the $Z^{\prime}$ boson
  and $Z$ boson as a function of $Z'$ mass. For comparison CMS data 
\cite{CMS:zprime} is
shown at different confidence levels of expectations. The solid
zig-zag curve denotes observed fluctuations about the median.}
\label{Fig:zpzratio} 
\end{figure}

The CMS collaboration \cite{CMS:zprime} has also measured the cross
section ratio ${{\sigma(pp\to Z'\to l^{\pm}l^{\mp} X)}/{\sigma(pp\to
    Z\to l^{\pm}l^{\mp} X)}}$ as shown in Fig. \ref{Fig:zpzratio} as
function of the $Z'$ mass $M_{Z'}$. Our prediction using $l=e$ is shown
by almost a slanted linear curve with falling value of the ratio with
increasing value of $M_{Z'}$. From this  Fig.\ref{Fig:zpzratio} we find
that in our model the lower limit of the $Z'$ mass is predicted to be
 $M_{Z'}\sim 2.8$ TeV.\\

\begin{figure}
\includegraphics[width=8cm,height=5cm,angle=0]{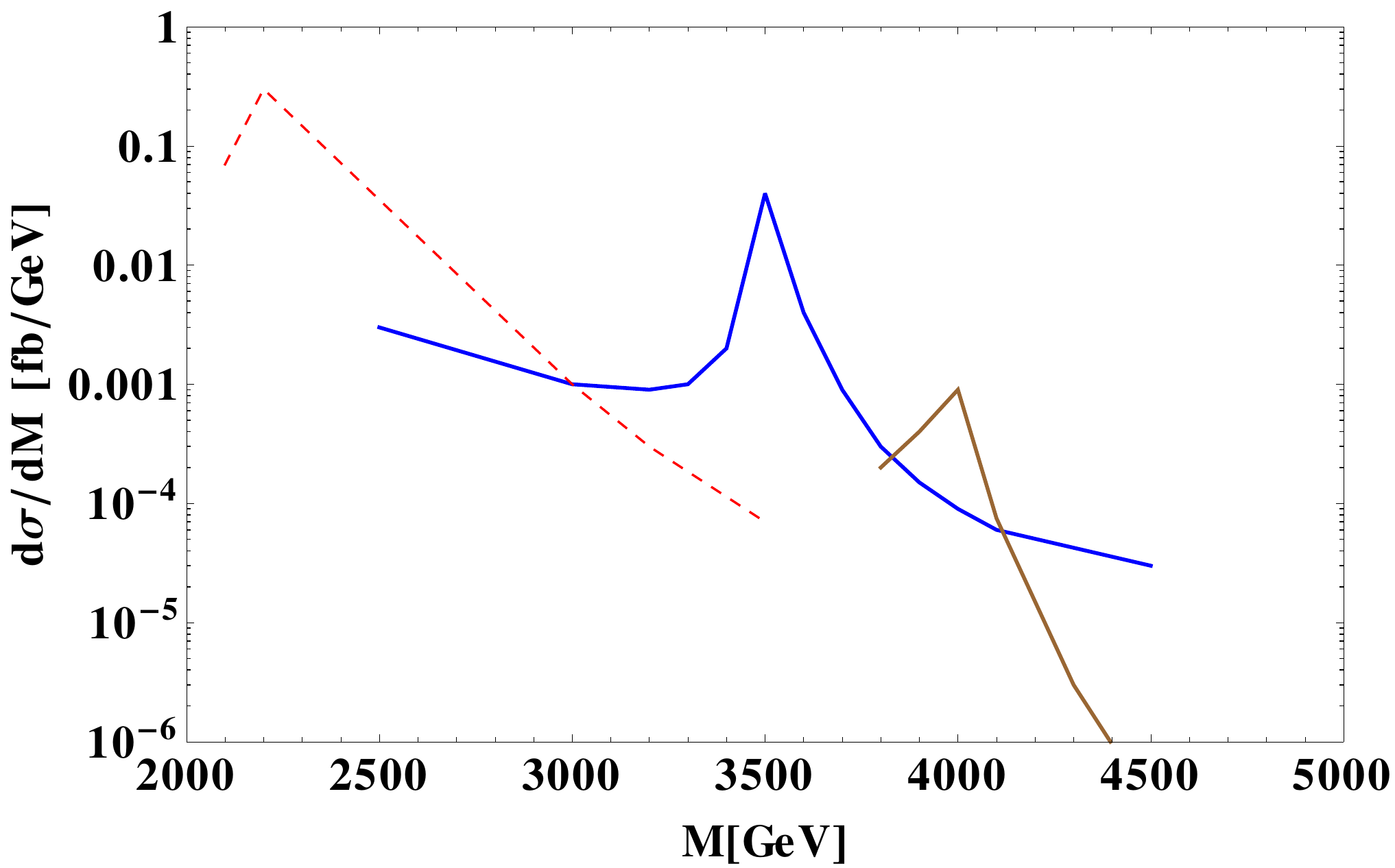}
\caption{Invariant mass distribution of $W$ pairs in $pp\to
  W^{+}W^{-}+ X$ at the LHC with $\sqrt S= 14$ TeV.}
\label{Fig:zpzdist} 
\end{figure}

The potential of LHC to discover $Z-Z^{\prime}$ mixing effects in the
process $pp\rightarrow W^{+}W^{-} + X$ has been investigated
\cite{Andreev}. In this paper we estimate the variation of
differential cross section with respect to the invariant mass $M$ of the produced
$W^+W^-$ pair for different $Z-Z^{\prime}$ mixings.
The corresponding differential cross section for the process $pp\rightarrow W^{+}W^{-} + X$ averaged over quark colors can be
obtained by \cite{Andreev}\\
\bea
\frac{d\hat{\sigma}^{Z^{\prime}}}{d{cos\theta}}&=&\frac{\pi\alpha^2{cot^2\theta_w}}{48}
\beta_{w}^3(v_{2,f}^2 \nonumber\\
&&+a_{2,f}^2)sin^2{\phi}\frac{\hat{s}}{(\hat{s}-M_{2}^2)^2+M_{2}^2\Gamma_{2}^2}
\nonumber\\
&&\times(\frac{\hat{s}^2}{M_w^4}sin^2\theta \nonumber\\
&&+4\frac{\hat{s}}{M_w^2}(4-sin^2\theta)+12sin^2\theta)
\eea

In our model for  invariant mass $2.3$ TeV, the mixings are 
$1.2\times10^{-3}$, $0.9\times10^{-3}$, and $0.7\times10^{-3}$ at
invariant mass values of  $2.3$ TeV, $3.5$ TeV, and  $4$ TeV,
respectively.

 The distribution curves are shown  in fig.(\ref{Fig:zpzdist}) for
 these three sets of values. Our predicted results give the value of
 $\frac{d\sigma}{dM} = 0.52$ ({fb}/{GeV}) for $M_{Z'}=3.5$ TeV. The
 predicted values of the peak positions at $2.3$ TeV and $4$ TeV can
 also provide a test of the models if such a $Z'$ is present in nature.  

\section{SUMMARY AND CONCLUSION}

In summary we find that the two $SO(10)$ models proposed recently to
predict TeV scale $Z'$ and RH neutrinos with charged lepton flavor
violating branching ratios only few to four orders smaller than the
experimental upper bounds make dominant
contributions to
neutrinoless double beta decay mediated by light sterile neutrino mass of the
first or the second generation. Although only the first generation sterile neutrino was shown
to mediate the double beta decay process in ref.\cite{bpnmkp:2015}, here we
have found that each of the first or the second generation sterile neutrino mass 
is allowed to be in the range of $\sim 1-10$ GeV and has the
capability to 
make dominant contributions to double beta decay while mediating
displaced vertices for like-sign dilepton signals of the type $eejj$,
$e\mu jj$,
and $\mu\mu jj$ outside the LHC detectors with drastically suppressed standard model
backgrounds and without missing energy. The predicted values of heavy
light mixings in Model-I and Model-II of non-SUSY $SO(10)$ theory are
found to overlap with regions of the parameter space identified in the
 model-independent approach under different cut conditions.    
For the $eejj$ process to be
visible with the heavy-light mixings compatible with double-beta decay
bound, a low momentum cut of the second electron is
necessary. Significant number of events in the $e\tau jj$ and $\mu \mu
jj$ channels are predicted with luminousity ${\cal L}= (50-300)$
fb$^{-1}$ which will either testify or falsify these models. 
 With the first generation sterile neutrino
mediating the double beta decay and the displaced vertices for $eejj$
and $\mu\mu jj$ events, 
 both the $SO(10)$ models explain the right
order of the baryon asymmetry of the universe through resonant
leptogenesis mediated by the heavy 
quasi-degenerate pair of the second and third generation sterile
neutrinos having mass $\sim 500$ GeV.
In the second alternative scenario  where dominant double beta decay
and  di-electron and 
 di-muon production through
 displaced vertices are mediated by the second sterile neutrino mass
 of  $\sim 10$ GeV,
 the resonant leptogenesis to explain the baryon asymmetry is
implemented by the quasi-degenerate pair of heavy sterile neutrinos of mass
$\sim 500$ GeV belonging to the first and the third generations. 
 These predictions with displaced vertices are
 easier to verify as the number of signal events are produced outside
 the detectors having  suppressed back grounds in the $l^{\pm}l^{\pm}
 jj$ channels without missing energy.\\

In this work we have also shown how the dominant double beta decay,
resonant leptogenesis, observable like-sign dilepton production via
displaced vertices along with non-unitarity effects and experimentally
accessible proton lifetime are possible in the two models in the
presence of NH pattern of light neutrino masses consistent with the recent
cosmological bound.

We have discussed two aspects of resonant $Z'$ production at the LHC
in the lepton pair $l^+l^-$ production and $W^+W^-$ pair production channels. 
While in the former case detection of $Z'$ with mass $M_{Z'}\ge 2.5$ TeV
would require beam luminosity $>1000$ fb$^{-1}$, in the latter case
it may easier to identify the peak structure even for reasonable
luminosities. At ILC the $Z'$ boson would manifests most prominently
by direct detection of its resonant peak at the predicted mass value. 
Finally we conclude that the two TeV scale $Z'$ models proposed recently
with Type-II see-saw dominance are quite effective in predicting
like-sign dilepton events via displaced vertices while explaining
neutrino oscillation data and baryon asymmetry of the universe with
predictions of significant leptonic non-unitarity effects, and LFV and LNV decays
accessible to ongoing searches.  Earlier it was noted that the proton
lifetime predictions are also accessible to experimental
searches.These models have high degree of falsifiability through a
number of experimental observables they predict.

\begin{center}           
      {\bf ACKNOWLEDGMENT }
\end{center}
M. K. P. thanks the Science and Engineering Research Board, Department
of Science and Technology, Government of India for grant of research
project SB/S2/HEP-011/2013. B.P.N. thanks  Siksha 'O' Anusandhan 
University for a research fellowship.\\

\end{document}